# Structural phase diagram and magnetic properties of Sc-substituted rare earth ferrites $R_{1-x}Sc_xFeO_3$ (R=Lu, Yb, Er, and Ho)

Jason White[1], Kishan Sinha[1], Xiaoshan Xu[1,2]

[1]*Department of Physics and Astronomy, University of Nebraska, Lincoln, NE 68588, USA*

[2]*Nebraska Center for Materials and Nanoscience, University of Nebraska, Lincoln, NE 68588, USA*

**Abstract**

We have studied the structural stability of Sc-substituted rare earth ($R$) ferrites $R_{1-x}Sc_xFeO_3$, and constructed a structural phase diagram for different $R$ and $x$. While $R$FeO$_3$ and ScFeO$_3$ adopt the orthorhombic and the bixbyite structure respectively, the substituted compound $R_{1-x}Sc_xFeO_3$ may be stable in a different structure. Specifically, for $R_{0.5}Sc_{0.5}FeO_3$, the hexagonal structure can be stable for small $R$, such as Lu and Yb, while the garnet structure is stable for larger $R$, such as Er and Ho. The formation of garnet structure of the $R_{0.5}Sc_{0.5}FeO_3$ compounds which requires that Sc occupies both the rare earth and the Fe sites, is corroborated by their magnetic properties.

## Introduction

Rare earth ferrites include insulating materials of magnetic orders significantly above room temperature, making them suitable for various applications. For example, orthoferrites, or orthorhombic $R$FeO$_3$ ($R$: rare earth), are weakly ferromagnetic based on an antiferromagnetic order below 600-700 K, which combined with their transparency, are great for magneto optical applications.[1–3] Rare earth iron garnet, or $R_3$Fe$_5$O$_{12}$, are unique insulating materials of ferrimagnetic order below about 550 K, which are critical for spintronics applications[4,5].

Hexagonal rare earth ferrites (h-$R$FeO$_3$, $R$: Er-Lu) exhibit coexisting ferroelectric and magnetic orders[6,7], making them multiferroics with application potential in novel sensors, actuators, and energy-efficient information storage and processing. The stable structure of $R$FeO$_3$ is orthorhombic [Fig. 1(a)][2,3], which is the origin of the name orthoferrite (o-$R$FeO$_3$). The hexagonal $R$FeO$_3$ (h-$R$FeO$_3$) [Fig. 1(b)] are unstable due to the expanded lattice compared with that of o-$R$FeO$_3$. Typically, h-$R$FeO$_3$ can be stabilized in epitaxial films with the aid of interfacial energy with the substrates[6,8–15]. Recently, it was demonstrated that the stability of the hexagonal structure of h-LuFeO$_3$ can be enhanced by doping Sc on the Lu site[16–18]. ScFeO$_3$ is unstable in the orthorhombic structure due to the small size difference between Sc and Fe compared with that between rare earth and Fe. Instead, it adopts a bixbyite structure [Fig. 1(c)] containing many open spaces.[19–21] Therefore, doping Sc on the $R$ site of h-$R$FeO$_3$ reduces the stability of the orthorhombic structure, which affects the competition between the orthorhombic and the hexagonal phases, leading to the enhanced stability of the hexagonal phase. Indeed, it was shown that by doping Sc, h-Lu$_{0.5}$Sc$_{0.5}$FeO$_3$ can be synthesized in bulk powder and even single-crystalline form[16–18].

The enhanced stability in h-Lu$_{0.5}$Sc$_{0.5}$FeO$_3$ has made the investigation of its properties more efficient.[22] On the other hand, the effect of Sc doping on the stability of other members of h-$R$FeO$_3$, as well as whether the stability can introduce the new members to the h-$R$FeO$_3$ family, have not been reported. In this work, we studied the effect of Sc doping in rare-earth ferrite compounds $R_{1-x}$Sc$_x$FeO$_3$, where $R$=Ho, Er, and Yb. The results show that h-Yb$_{0.5}$Sc$_{0.5}$FeO$_3$ can be stabilized in polycrystalline powders, but with impurity phases; for larger $R$, h-$R_{0.5}$Sc$_{0.5}$FeO$_3$ crystalize in a pure garnet phase.

## Experimental

Powder samples were synthesized using solid-state reactions starting from $R_2$O$_3$, Fe$_2$O$_3$, and Sc$_2$O$_3$ powders purchased from Alfa Aesar. The samples were mixed stoichiometrically using a mortar and a pestle before being calcinated at 900 °C for 12 hours in a tube furnace with a constant oxygen flow. The samples were then crushed and ground again before being pressed into pellets and sintered at 1200 °C for 18 hours. The orthoferrite $R$FeO$_3$ pellets were made first by reacting $R_2$O$_3$ and Fe$_2$O$_3$ using the above method. The ScFeO$_3$ pellets were made similarly by reacting Sc$_2$O$_3$ and Fe$_2$O$_3$ powders. The $R_{1-x}$Sc$_x$FeO$_3$ compounds were then made by reacting $R$FeO$_3$ and ScFeO$_3$ powders according to the corresponding stoichiometry. Ho$_3$Fe$_5$O$_{12}$ garnet samples were also made from Ho$_2$O$_3$ and Fe$_2$O$_3$. X-ray diffraction measurements were carried out using a Rigaku D/Max-B diffractometer with Co-K radiation ($\lambda$ = 1.789 Å). The temperature and magnetic-field dependence of the magnetization were measured using a vibrating sample magnetometer (VSM).

**Results and discussion**

*Structural properties*

To investigate the effect of Sc doping on the stability of h-$R$FeO$_3$, we synthesized polycrystalline $R_{0.5}$Sc$_{0.5}$FeO$_3$ compound for $R$ = Yb, Er, and Ho. As shown in Fig. 2, For $R$=Yb, hexagonal phase was observed in the x-ray diffraction spectra; in addition, the orthorhombic phase and a garnet phase were also observed. When $R$ = Er and Ho, no hexagonal, orthorhombic, or bixbyite structures were observed. The solid-state reaction appears to result in a pure garnet structure phase, which can be seen by the comparison between the x-ray diffraction spectra with that of Ho$_3$Fe$_5$O$_{12}$, as shown in Fig. 2(a).

The observation of pure garnet phase in both Ho$_{0.5}$Sc$_{0.5}$FeO$_3$ and Er$_{0.5}$Sc$_{0.5}$FeO$_3$ suggests the stability of the garnet structure. To further investigate the effect of Sc doping on the stability of this structure, we synthesized Ho$_{1-x}$Sc$_x$FeO$_3$ of different $x$ between 0.1 and 0.9. As shown in Fig. 3, when $x$ is small, orthorhombic structure (stable for $R$FeO$_3$)[2,3] dominates, while when $x$ is large, the bixbyite structure (stable for ScFeO$_3$)[19,20] dominates, as expected. On the other hand, the garnet phase always exists for 0.1≤$x$≤0.9. When $x$=0.5, both bixbyite and orthorhombic phases disappear, leaving a pure garnet phase.

We constructed the structural phase diagram (Fig. 4) of $R_{1-x}$Sc$_x$FeO$_3$ using the data from this and the previous work[19]. Two variables in the phase diagram are $x$ and $R$, representing the Sc/rare earth stoichiometry and the species of the rare earth respectively. According to Fig. 4, the hexagonal phase can be stabilized in $R_{1-x}$Sc$_x$FeO$_3$ of smaller rare earth (Yb and Lu). In particular, pure hexagonal Lu$_{1-x}$Sc$_x$FeO$_3$ (0.4≤$x$≤0.6) phase can be obtained. $R_{1-x}$Sc$_x$FeO$_3$ of larger rare earth (Er and Ho) favor the garnet phase instead.

The appearance of the garnet phase is surprising since all the competing structural phases considered previously, including the bixbyite, the orthorhombic, and the hexagonal structural phases have the $R$FeO$_3$ chemical formula, while the garnet has a $R_3$Fe$_5$O$_{12}$ chemical formula. This puzzle may be resolved if we rewrite $R_{0.5}$Sc$_{0.5}$FeO$_3$ as ($R_2$Sc)(Fe$_4$Sc)O$_{12}$, assuming that Sc can occupy both the rare earth site and the Fe site.

To verify this proposal that Sc can occupy both the rare earth site and the Fe site, we carried out crystal structure refinement of $R_{0.5}$Sc$_{0.5}$FeO$_3$ and Ho$_3$Fe$_5$O$_{12}$ by fitting the diffraction spectra with the FullProf program[23,24] using the garnet structure which has a cubic symmetry; the comparison between the observation and the fit are displayed in Fig. 2(b-d); the refinement parameters are shown in Table I. The refinement on Ho$_3$Fe$_5$O$_{12}$ results in a lattice constant 12.365 Å, in good agreement with the previous work[25,26]. The closeness of expected and weight-profile $R$-factor ($R_{exp}$ and $R_{wp}$) suggests fair agreement within the experimental uncertainty. For Er$_{0.5}$Sc$_{0.5}$FeO$_3$ and Ho$_{0.5}$Sc$_{0.5}$FeO$_3$, we assumed that the Sc may substitute $R$ on the "$c$" site and Fe on the "$a$" site respectively because Sc is too large to fit in the "$d$" site[26,27]. As shown in Table I, in Ho$_{0.5}$Sc$_{0.5}$FeO$_3$, the best fit corresponds to an occupancy of 0.74 and 0.26 ± 0.1 for Ho and Fe on the "$c$" sites, and 0.44 and 0.56 ± 0.05 for Fe and Sc on the "$a$" sites respectively; in Er$_{0.5}$Sc$_{0.5}$FeO$_3$, the occupancy is 0.76 and 0.24 ± 0.1 for Er and Sc on the "$c$" sites respectively, and 0.5 and 0.5 ± 0.05 for Fe and Sc on the "$a$" sites respectively. In both Ho$_{0.5}$Sc$_{0.5}$FeO$_3$ and Er$_{0.5}$Sc$_{0.5}$FeO$_3$, within the experimental uncertainty, the site occupancy agrees with the proposed chemical formula $(R_2Sc)(Fe_4Sc)O_{12}$, where Sc substitutes 1/3 of the $R$ atoms on the "$c$" sites and ½ of the Fe atoms on the "$a$" sites on average. Therefore, the result in Table I provides direct evidence for the garnet structure and chemical composition of $Ho_{0.5}Sc_{0.5}FeO_3$ and $Er_{0.5}Sc_{0.5}FeO_3$.

To understand why Sc can occupy both the rare earth sites and the Fe sites, we surveyed the effective ionic radii of trivalent transition metal and rare earth elements.[28] As shown in Fig. 5(a), the $R^{3+}$ ions are larger than the transition metal ions, which is why $R^{3+}$ tend to have a coordination number (CN) 8, i.e., surrounded by 8 oxygen ions. In contrast, the coordination numbers of trivalent transition metal ions are smaller; typically, CN=6. On the other hand, the coordination number of $Sc^{3+}$ can be both 6 and 8. When CN=8, the radius of $Sc^{3+}$ is close to that of $R^{3+}$; when CN=6, the radius of $Sc^{3+}$ is close to that of trivalent transition metal ions. Therefore, it is possible that $Sc^{3+}$ ion can occupy both the rare earth site and the Fe site.

As shown in Fig. 5(b), the lattice constant of the $R_{0.5}Sc_{0.5}FeO_3$ garnet structure ($a_g$) decreases when the $R$ changes from Ho to Yb, due to a decreasing ionic radius from $Ho^{3+}$ to $Yb^{3+}$. In comparison to the lattice constants of $R_3Fe_5O_{12}$ (from the previous work[26]), the $a_g$ values of $R_{0.5}Sc_{0.5}FeO_3$ garnets are systematically smaller. Since replacing $R$ atoms with Sc is expected to make the lattice constant smaller because $Sc^{3+}$ is smaller than $Ho^{3+}$ [Fig. 5(a)], this observation corroborates the substitution of Fe with Sc, which is expected to expand the lattice constant since $Sc^{3+}$ is larger than $Fe^{3+}$.

We also analyzed the lattice constants of the garnet ($a_g$) and the bixbyite ($a_b$) structural phases for the $Ho_{1-x}Sc_xFeO_3$ samples; both structures have a cubic symmetry. As shown in Fig. 5(c), when $x$ is small, the lattice constant of the garnet phase $a_g$ is larger than that of $Ho_3Fe_5O_{12}$, indicating Sc is going into the Fe sites first. When $x$ increases, corresponding to more Sc, $a_g$ decreases monotonically, suggesting that Sc atoms start to occupy more on the Ho sites. At $x = 0.8$, a rapid drop of $a_g$ is observed, which is accompanied by the appearance of the bixbyite structural phase [see Fig. 5(c)]. The lattice constant of the garnet phase $a_g$ drops to values close to that of $Ho_3Fe_5O_{12}$. At the same time, the lattice constant of the bixbyite structural phase $a_b$ is also smaller than that of $ScFeO_3$. These observations indicate that when $x$ is too large, the garnet structure for $Ho_{1-x}Sc_xFeO_3$ is unstable against the decomposition into garnet $Ho_3Fe_5O_{12}$ and Fe-rich $ScFeO_3$.

The lattice constants for the hexagonal Sc-substitute $YbFeO_3$ phase was also analyzed, resulting $a_h = 5.83$ Å and $c_h = 11.69$ Å. These numbers are slightly smaller than the lattice constants of h-$Lu_{0.5}Sc_{0.5}FeO_3$ ($a_h = 5.86$ Å and $c_h = 11.71$ Å), but closer to those of $Lu_{0.4}Sc_{0.6}FeO_3$ ($a_h = 5.83$ Å and $c_h = 11.705$ Å) reported previously[17]. This observation may indicate that more Sc doping is needed to stabilize hexagonal $YbFeO_3$ compared with that of hexagonal $LuFeO_3$.

*Magnetic properties*

Next, we turn to the magnetic properties of $Ho_{0.5}Sc_{0.5}FeO_3$ and $Er_{0.5}Sc_{0.5}FeO_3$ garnets, as well as that of $Ho_3Fe_5O_{12}$ as a reference. Figure 6(a) shows the temperature dependence of magnetization $M(T)$ measured in a 700 Oe magnetic field, where $T$ is temperature. All three compounds share the typical features of rare earth iron garnets: an above-room-temperature magnetic ordering temperature (Curie temperature $T_C$), and a maximum and a downturn of magnetization at low temperature which comes from the compensation between the Fe and the rare earth magnetic moments that causes a minimum of magnetization at the compensation temperature ($T_{cmp}$).[4] From Fig. 6(a), we found $T_C \approx 600$ K and $T_{cmp} \approx 130$ K for $Ho_3Fe_5O_{12}$, $T_C \approx 500$ K and $T_{cmp} \approx 63$ K for $Ho_{0.5}Sc_{0.5}FeO_3$ garnet, and $T_C \approx 500$ K and $T_{cmp} < 50$ K for $Er_{0.5}Sc_{0.5}FeO_3$ garnet. By comparing the results of $Ho_{0.5}Sc_{0.5}FeO_3$ garnet and $Ho_3Fe_5O_{12}$, $Er_{0.5}Sc_{0.5}FeO_3$ garnet and $Er_3Fe_5O_{12}$ (data from the previous work $T_C \approx 560$ K and $T_{cmp} \approx 80$ K)[4], one observes that both $T_C$ and $T_{cmp}$ are reduced after substituting Sc in Ho and Er iron garnet.

Figure 7(a) and (b) shows the dependence of the magnetization on magnetic field $M(H)$ for various temperatures, where $H$ is the external magnetic field. For both $Ho_{0.5}Sc_{0.5}FeO_3$ and $Er_{0.5}Sc_{0.5}FeO_3$ garnets, the coercive field appears to be smaller than 500 Oe. At high field, the magnetization rises linearly with $H$, as observed in other iron garnets.[4] By fitting the linear part of data in Fig. 7(a) and (b) ($|H| > 5$ kOe) using the formula $M(H) = M_{int} + (dM/dH) H$, the slope $dM/dH$ and the intercept $M_{int}$ can be found, as displayed in Fig. 7(c) and (d). While $dM/dH$ decreases monotonically with temperature, $M_{int}$ shows a maximum and a downturn at low temperature, similar to the trend of $M(T)$ displayed in Fig. 6(a).

To understand the temperature and magnetic-field dependence of magnetization of $Ho_{0.5}Sc_{0.5}FeO_3$ and $Er_{0.5}Sc_{0.5}FeO_3$ garnets, and the effect of Sc-doping, we summarize the basics of crystal and magnetic structures of $R_3Fe_5O_{12}$. Displayed in Fig. 6(b) is 1/8 of the unit cell of $R_3Fe_5O_{12}$, containing two "$a$" sites, three "$d$" sites, and three "$c$" sites which are occupied by the metal atoms. The two "$a$" sites (octahedra local environment with six-fold oxygen coordination), and the three "$d$" sites (tetrahedral local environment with four-fold oxygen coordination) are occupied by Fe atoms, while the three "$c$" sites (dodecahedral local environment with eight-fold oxygen coordination) are occupied by the rare earth atoms.[29] When only the nearest neighbors are considered, the Fe magnetic moments on the "$a$" sites couple to those on the "$d$" sites antiferromagnetically to form a ferrimagnetic order below $T_C$.[4] Because there are more "$d$" sites than "$a$" sites, the net magnetic moment is parallel to those on the "$d$" sites. The coupling between the rare earth and the Fe magnetic moments is much weaker than that between the Fe sites. For not-so-low temperature (above a few K), the rare earth sites can be approximately described in the mean field theory as individual paramagnetic moments magnetized by the exchange field from the Fe sites ($H_{R-Fe}$) which is opposite to the Fe moments.[5]

To analyze the result in Fig. 5 and Fig. 6, we write the total magnetization as

$$M_{total} = M_{Fe} + M_R, \quad (1)$$

where $M_{Fe}$ and $M_R$ are magnetization of Fe and rare earth ions respectively.

Treating the contribution of rare earth ions as individual paramagnetic ions, one has

$$M_R = N_R\, \mu_R\, B_r(y), \quad (2)$$

where $y \equiv \mu_0\, \mu_R(H + H_{R-Fe})/(k_B T)$, $N_R$ and $\mu_R$ are the density and the magnetic moment of the rare earth sites, $\mu_0$ and $k_B$ are the vacuum permeability and the Boltzmann constant, $H$ and $T$ are the external magnetic field and temperature respectively, and $B_r(y)$ is the Brillouin function.

Since $B_r(y) \approx y/3$ for $y \ll 1$, if $\mu_0 \mu_R(H+ H_{R\text{-}Fe})/(k_BT) \ll 1$, Eq. (2) can be simplified as

$$M_R = N_R \mu_0 \mu_R^2 (H+ H_{R\text{-}Fe})/(3k_BT) \qquad (3)$$

Assuming $H_{R\text{-}Fe} = -\Gamma M_{Fe}$, where $\Gamma > 0$, $\qquad (4)$

$$M_{total} = M_{Fe} + N_R \mu_0 \mu_R^2 (H-\Gamma M_{Fe})/(3k_BT)$$

$$= M_{Fe} [1-N_R \mu_0 \mu_R^2\Gamma/(3k_BT)] + N_R \mu_0 \mu_R^2 H/(3k_BT). \qquad (5)$$

When $1-N_R \mu_0 \mu_R^2\Gamma/(3k_BT) = 0$, the total magnetization $M_{total}$ has a minimum, corresponding to the compensation temperature. In addition, if $1-N_R \mu_0 \mu_R^2\Gamma/(3k_BT) < 0$, to maximize magnetization or to minimize the total energy, $M_{Fe}$ becomes antiparallel to the external field.

At high enough $H$, $M_{Fe}$ is expected to saturate. According to the second term in Eq. (5), the total magnetization $M_{total}$ is then linearly proportional to $H$, as also observed in Fig. 7(a) and (b). In addition, the slope $dM/dH$ is inversely proportional to $T$ at high field according to Eq. (5), as also observed in Fig. 7(c) and (d). The intercept for the linear part of the high-field magnetization corresponds to the first term of Eq. (5), which is expected to reflect the magnetization compensation, i.e., a maximum and down turn at low temperature of the magnetization, as observed in Fig. 7(c) and (d).

The reduction of $T_{cmp}$ by replacing rare earth with Sc is expected from Eq. (5), which leads to $T_{cmp} = N_R \mu_0 \mu_R^2\Gamma/(3k_B) \propto N_R\Gamma$. Since $Sc^{3+}$ has no magnetic moment, the density of magnetic $R^{3+}$ ($N_R$) is reduced when they are replaced by $Sc^{3+}$. Higher magnetization of rare earth ions is then needed to compensate the Fe magnetization, which requires lower temperature. Therefore, replacing magnetic $R^{3+}$ with $Sc^{3+}$ is expected to lower $T_{cmp}$, as observed in Fig. (5) as well as previously in Y-substituted (on Gd sites) $Gd_3Fe_5O_{12}$.[30] A quantitative estimation of the reduction of $T_{cmp}$ can be carried out using the site occupancy found in Table I. Clearly, $N_R$ is proportional to the site occupancy of the rare earth atoms; $\Gamma$ is proportional to coordination number of rare earth atoms in the Fe environment, which is proportional to the Fe site occupancy. Therefore, using the data in Table I, we estimate that $T_{cmp}(Ho_{0.5}Sc_{0.5}FeO_3)/T_{cmp}(Ho_3Fe_5O_{12}) = 0.57 \pm 0.08$ and $T_{cmp}(Er_{0.5}Sc_{0.5}FeO_3)/T_{cmp}(Er_3Fe_5O_{12}) = 0.59 \pm 0.08$, which leads to the prediction of $T_{cmp}(Ho_{0.5}Sc_{0.5}FeO_3) = 75 \pm 10$ K and $T_{cmp}(Er_{0.5}Sc_{0.5}FeO_3) = 49 \pm 6$ K, consistent with the observation in Fig 6(a) where $T_{cmp}(Ho_{0.5}Sc_{0.5}FeO_3) \approx 63$ K and $T_{cmp}(Er_{0.5}Sc_{0.5}FeO_3) < 50$ K.

The reduction of $T_C$ by doping Sc can be understood considering that Sc atoms also occupy the Fe sites. The replacement of Fe with Sc disrupts the Fe-Fe exchange coupling for the ferrimagnetic order. Because $Sc^{3+}$ has no magnetic moment, replacing Fe with Sc leads to a reduction of the effective Fe-Fe coordination number, which is expected to lower the magnetic ordering temperature $T_C$[31], as also observed in Ga-substituted (on Fe sites) $Y_3Fe_5O_{12}$ and Al-substituted (on Fe sites) $Gd_3Fe_5O_{12}$.[4,32]

**Conclusion**

We have studied the stabilization of the hexagonal structural phase of $RFeO_3$ by partially replacing the rare earth elements with Sc to destabilize the competing orthorhombic structural phase. The results indicate that for smaller rare earth, such as Lu and Yb, the stability of the hexagonal phase is enhanced, leading to the existence of hexagonal $Lu_{0.5}Sc_{0.5}FeO_3$ and $Yb_{0.5}Sc_{0.5}FeO_3$. In contrast, for larger rare earth, such as Ho and Er, the destabilization of orthorhombic phase accompanies the stabilization of the garnet phase instead of the hexagonal phase. The effect of doping Sc leads to the reduction of both Curie and compensation temperatures of the garnet phase, which can be understood by the occupancy of Sc on both the rare earth and the Fe sites. These results suggest strong tunability of rare earth ferrites both in crystal structure and magnetic properties, which could be useful for future material design and engineering.

**Acknowledgement**

This work was primarily supported by the National Science Foundation (NSF), Division of Materials Research (DMR) under Grant No. DMR-1454618. The research was performed in part in the Nebraska Nanoscale Facility: National Nanotechnology Coordinated Infrastructure and the Nebraska Center for Materials and Nanoscience, which are supported by the National Science Foundation under Grant No. ECCS-1542182, and the Nebraska Research Initiative. Characterization analysis were also performed in part at the NanoEngineering Research Core Facility, University of Nebraska-Lincoln, which is partially funded from Nebraska Research Initiative funds.

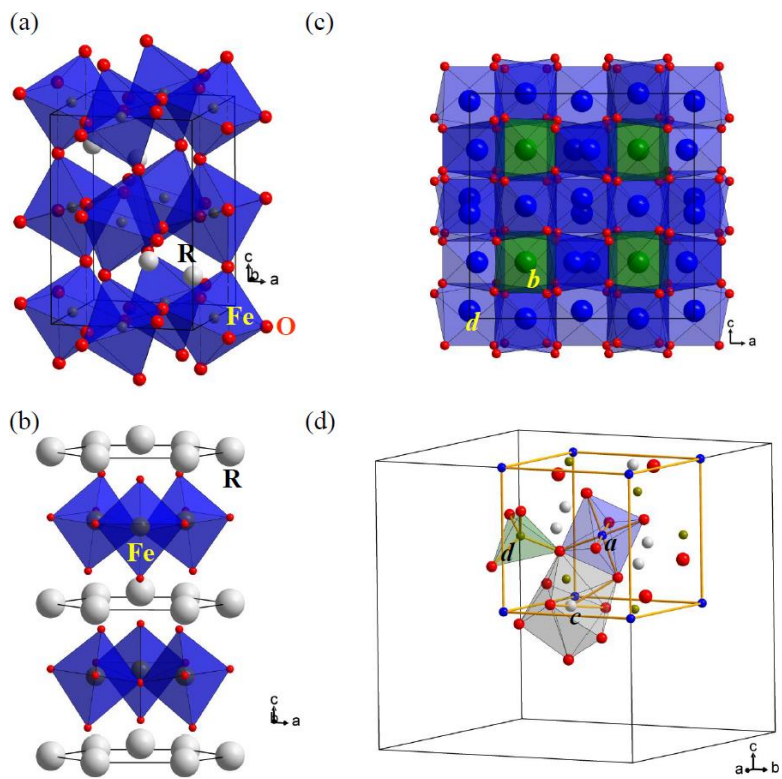

**Figure 1**. Crystal structures of (a) the orthorhombic and (b) the hexagonal $R$FeO$_3$ showing the unit cells. (c) Crystal structure of the bixbyite ScFeO$_3$, with two types of metal sites ("*b*" and "*d*") which are occupied by both Sc and Fe. (d) Crystal structure of the garnet $R_3$Fe$_5$O$_{12}$ showing 1/8 of the unit cell and three types of metal sites ("*a*", "*d*", and "*c*"). The "*a*" and "*d*" sites are occpupied by Fe; the "*c*" sites are occupied by rare earth.

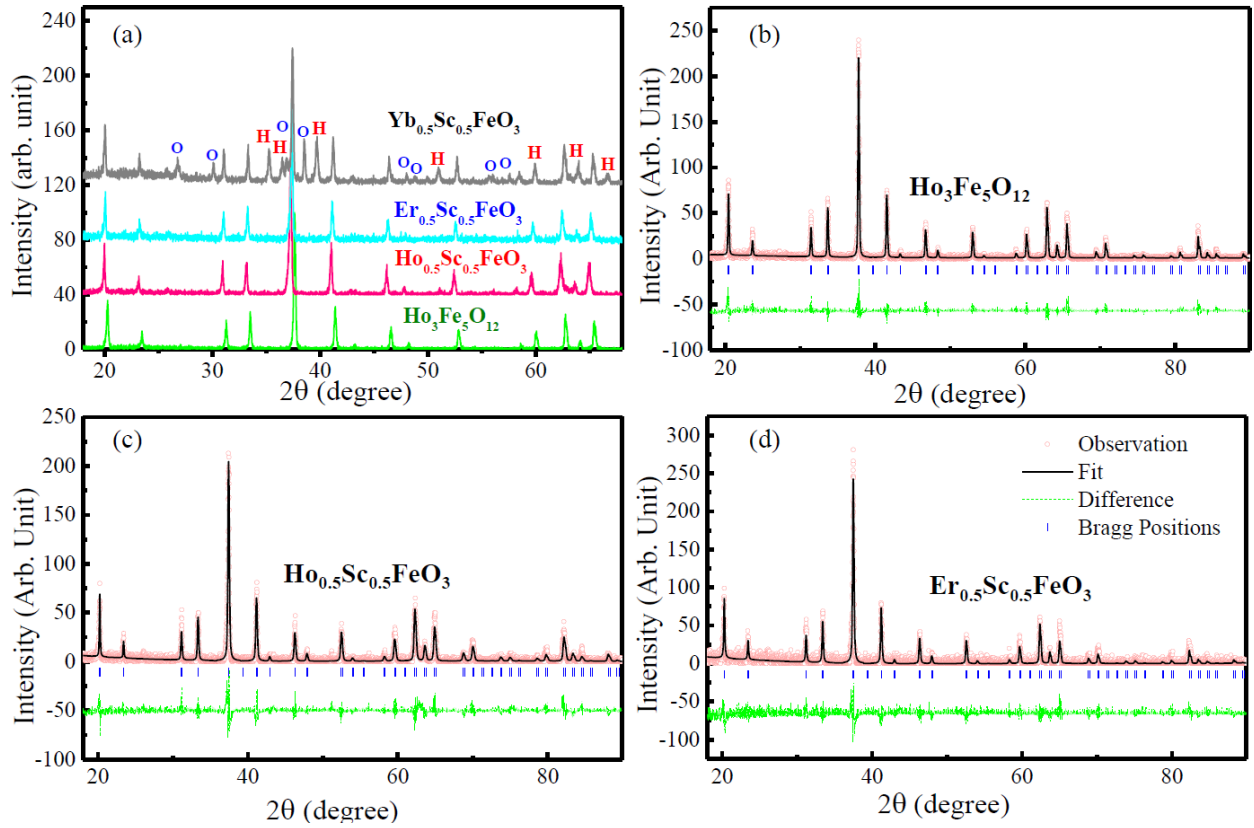

**Figure 2.** (a) Powder x-ray diffraction spectra for $R_{0.5}Sc_{0.5}FeO_3$ with different $R$, as well as $Ho_3Fe_5O_{12}$ as a reference. H and O stand for the hexagonal and orthorhombic structures respectively. (b)-(c) Structural refinement of $Ho_3Fe_5O_{12}$, $Ho_{0.5}Sc_{0.5}FeO_3$, and $Er_{0.5}Sc_{0.5}FeO_3$ respectively using the garnet structure (see text).

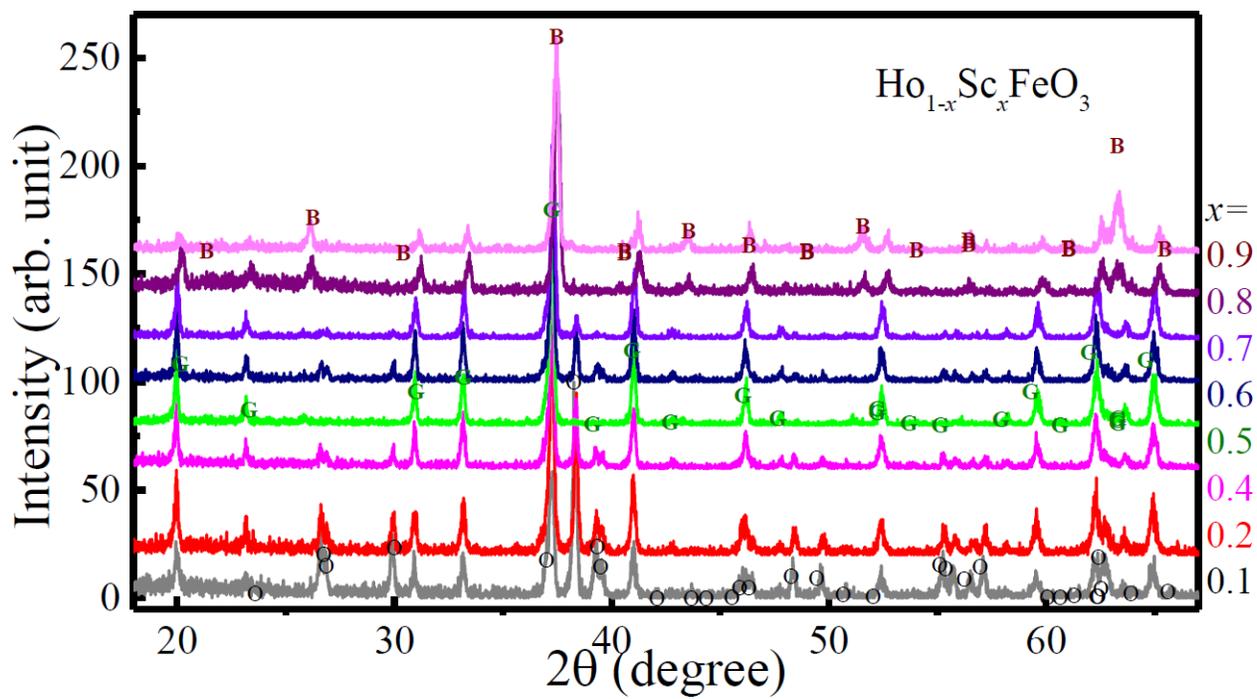

**Figure 3**. Powder x-ray diffraction spectra for $Ho_{1-x}Sc_xFeO_3$ with different *x*, where B, G, O represents bixbyite, garnet, and orthorhombic structures.

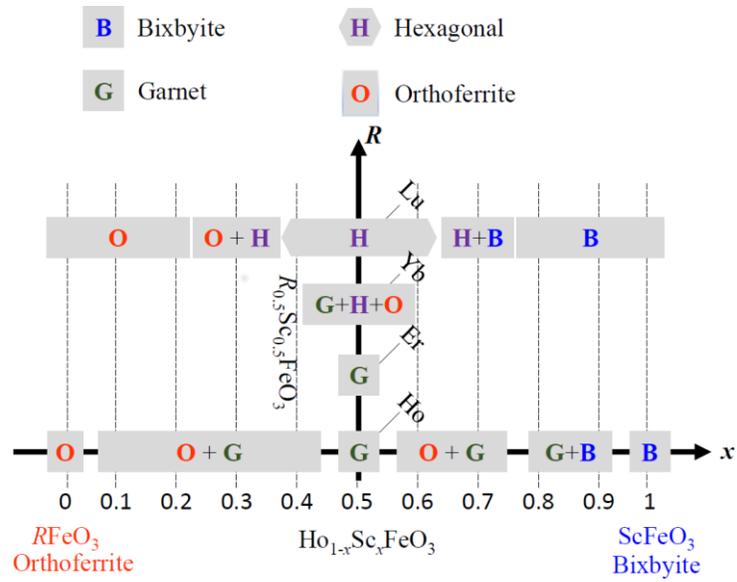

**Figure 4**. Structural phase diagram of Sc-substituted rare earth ferrites. The two dimensions are Sc/rare earth ratio $x$ and the rare earth species $R$.

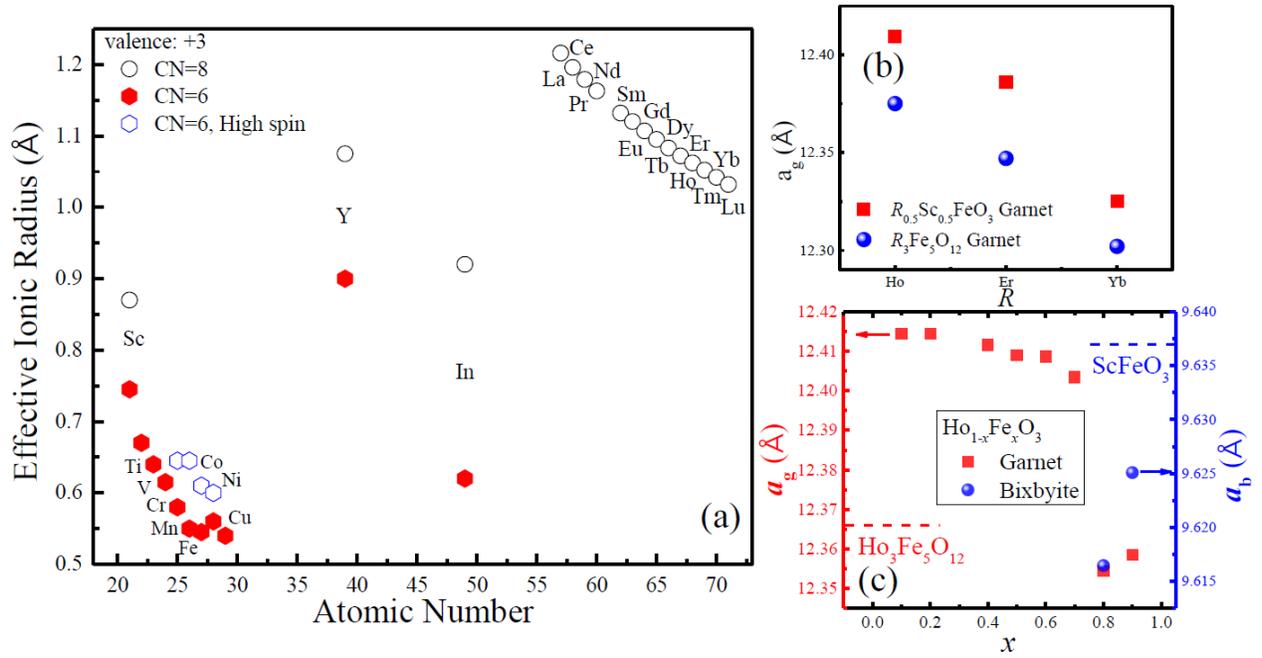

**Figure 5**. (a) The ionic radius as a function atomic number. The hexagons correspond to coordination number (CN) = 6, while the circles correspond to CN=8. (b) The lattice constants of the $R_{0.5}Sc_{0.5}FeO_3$ garnet (measured in this work) and $R_3Fe_5O_{12}$ garnet (from previous work[26]) for different $R$. (c) The lattice constants of the garnet and bixbyite structural phases in $Ho_{1-x}Sc_xFeO_3$ as a function of $x$. The lattice constants of $Ho_3Fe_5O_{12}$ and $ScFeO_3$ are indicated by the dashed lines.

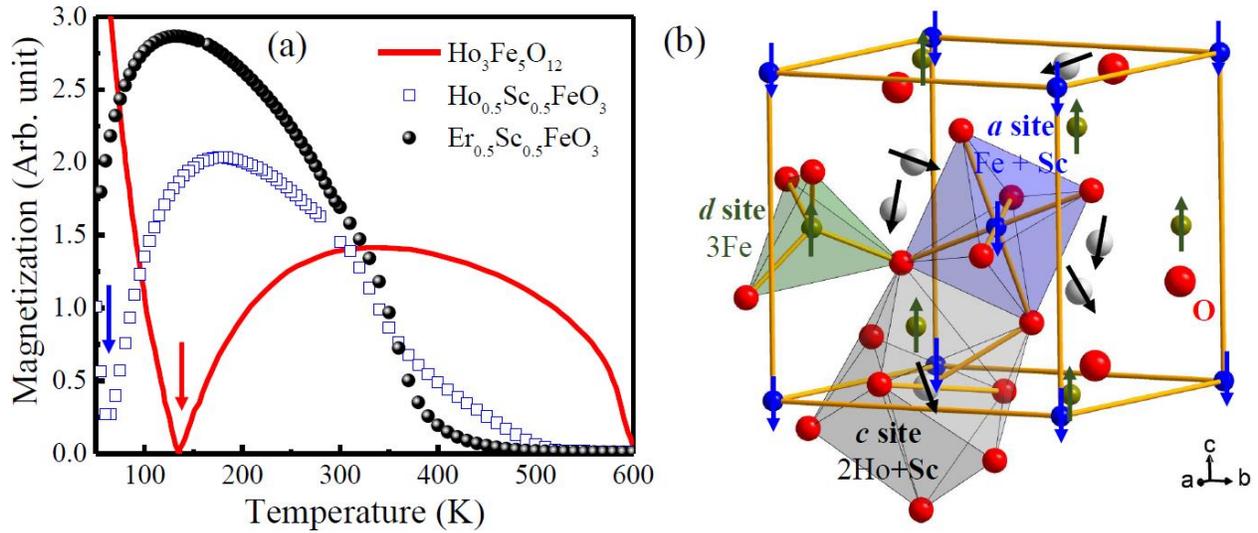

**Figure 6**. (a) Temperature dependence of magnetization of $Ho_{0.5}Sc_{0.5}FeO_3$, $Er_{0.5}Sc_{0.5}FeO_3$, and $Ho_3Fe_5O_{12}$ garnets measured in a 700 Oe magnetic field, respectively. The arrows indicate compensation temperatures. (b) Crystal structure and magnetic structure of iron garnets shown using 1/8 of the unit cell. The arrows indicate the magnetic moments on the metal sites. While the magnetic moments on the "*a*" sites couple to those on the "*d*" sites antiferromagnetically, the moments on the "*c*" sites are aligned by the exchange field from the "*a*" and "*d*" sites paramagnetically.

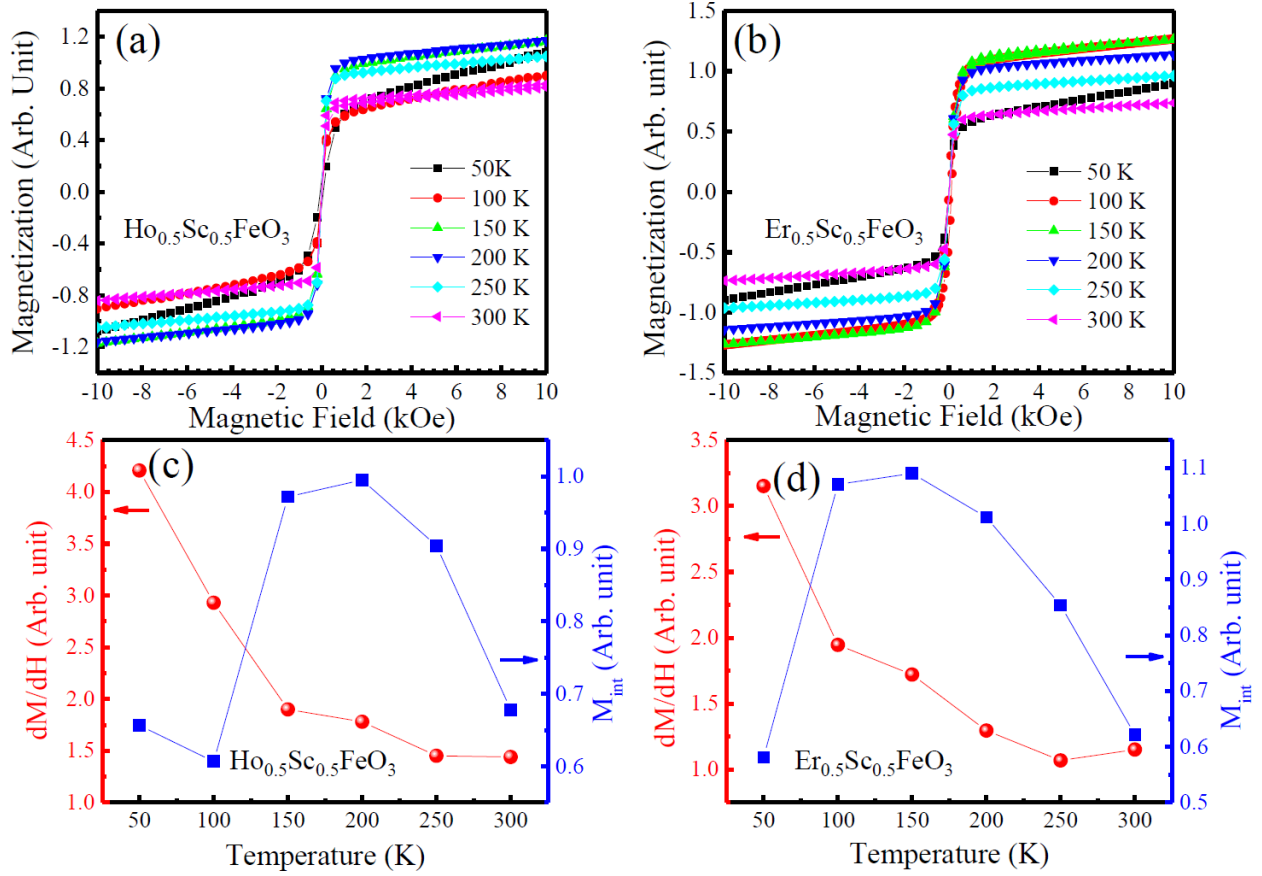

**Figure 7**. The magnetic-field dependence of the magnetization $M(H)$ of $Ho_{0.5}Sc_{0.5}FeO_3$ (a) and $Er_{0.5}Sc_{0.5}FeO_3$ (b) at various temperatures. (c) and (d) are the slope ($dM/dH$) and the intercept ($M_{int}$) of the high-field linear part of $M(H)$ in (a) and (b) respectively (see text).

**Table I**. Parameters found from structural refinement using the FullProf program.

| | | $Ho_3Fe_5O_{12}$ | $Ho_{0.5}Sc_{0.5}FeO_3$ | $Er_{0.5}Sc_{0.5}FeO_3$ |
|---|---|---|---|---|
| Lattice constant (Å) | | 12.365 | 12.478 | 12.458 |
| Fe1/Sc1 (16a) | Position | (0, 0, 0) | (0, 0, 0) | (0, 0, 0) |
| | Occupancy | 1 | 0.44/0.56 ± 0.05 | 0.5/0.5 ± 0.05 |
| Fe2 (24d) | Position | (3/8, 0, 1/4) | (3/8, 0, 1/4) | (3/8, 0, 1/4) |
| | Occupancy | 1 | 1 | 1 |
| R/Sc (24c) | Position | (1/8, 0, 1/4) | (1/8, 0, 1/4) | (1/8, 0, 1/4) |
| | Occupancy | 1 | 0.74/0.26 ± 0.1 | 0.76/0.24 ± 0.1 |
| O (96h) | Position | (0.1509, -0.0236, 0.0626) | (0.1471, -0.0292, 0.0692) | (0.1500, -0.0262, 0.0707) |
| | Occupancy | 1 | 1 | 1 |
| $\chi^2$ | | 2.16 | 1.66 | 3.50 |
| $R_{wp}$ (%) | | 22.0 | 29.5 | 39.1 |
| $R_{exp}$ (%) | | 20.9 | 22.93 | 20.9 |